\RequirePackage{fix-cm}
\documentclass[smallextended]{svjour3}       
\smartqed  
\usepackage{graphicx}
\usepackage{amsmath}
\usepackage{enumerate}
\usepackage{graphicx}
\usepackage{setspace}
\usepackage{filecontents}
\usepackage{amsmath,amssymb,multirow,epsfig} 
\usepackage{epsfig}
\usepackage{array}
\usepackage{psfrag}
\usepackage{pstricks,pst-node,pst-text,pst-3d,pst-plot}
\usepackage{trig}
\usepackage{amssymb}
\usepackage{anysize}
\usepackage{color}
\usepackage{cite}
\usepackage{tikz}
\usepackage{amsmath}
\begin{document}

\title{A ReRAM Physically Unclonable Function (ReRAM PUF)-based Approach to Enhance Authentication Security in Software Defined Wireless Networks \thanks{This work was partially supported by Arizona Board of Regents, grants number: 1003073 \& 1003074.}
}


\author{Fatemeh Afghah, Bertrand Cambou, Masih Abedini, Sherali Zeadally}


\institute{Fatemeh Afghah, Bertrand Cambou, Masih Abedini \at
              School of Informatics, Computing and Cyber Systems,
              Northern Arizona University, Flagstaff, USA 86011\\
              Tel.: +1928-523-5095\\
              \email{\{fatemeh.afghah, bertrand.cambou\}@nau.edu, masih@ieee.org}           
           \and
           Sherali Zeadally  \at
              College of Communication and Information,
              University of Kentucky, Lexington, KY 40506-0224\\
                \email{ szeadally@uky.edu}
}

\date{Received: date / Accepted: date}

\maketitle

\begin{abstract}
The exponentially increasing number of ubiquitous wireless devices connected to the Internet in Internet of Things (IoT) networks highlights the need for a new paradigm of data flow management in such large-scale networks under software defined wireless networking (SDWN). The limited power and computation capability available at IoT devices as well as the centralized management and decision making approach in SDWN introduce a whole new set of security threats to the networks. In particular, the authentication mechanism between the controllers and the forwarding devices in SDWNs is a key challenge from both secrecy and integrity aspects. Conventional authentication protocols based on public key infrastructure (PKI) are no longer sufficient for these networks considering the large-scale and heterogeneity nature of the networks as well as their deployment cost, and security vulnerabilities due to key distribution and storage.
We propose a novel security protocol based on physical unclonable functions (PUFs) known as hardware security primitives to enhance the authentication security in SDWNs. In this approach, digital PUFs are developed using the inherent randomness of the nanomaterials of Resistive Random Access Memory (ReRAM) that are embedded in most IoT devices to enable a secure authentication and access control in these networks. These PUFs are developed based on a novel approach of multi states, in which the natural drifts due to the physical variations in the environment are predicted to reduce the potential errors in challenge-response pairs of PUFs being tested in different situations. We also proposed a PUF-based PKI protocol to secure the controller in SDWNs. The performance of the developed ReRAM-based PUFs are evaluated in the experimental results. Moreover, the effectiveness of the proposed multi-state machine learning technique to predict the drifts of the PUFs' responses in different temperature and biased conditions is presented. 
\end{abstract}

\section{Introduction}
With the ever-growing number of devices connected to the Internet and the development of Internet of Things (IoT) networks, security of such large-scale heterogeneous networks has become a key challenge in cyber physical systems. Software defined networking (SDN) provides several benefits toward control and management of IoT networks.
In traditional infrastructure-based networks, the control and data planes are tightly coupled together to process packets according to the protocols defined individually in the control plane. The tight coupling of the data and control planes hinders flexibility and performance of such networks. In these networks, whenever the network administrators need to change or update a parameter of the protocols, they may need to re-configure all related devices (i.e. routers, switches, and firewalls) throughout the network. Depending on the size of the network, it can be a burdensome and time-consuming process.

The recently developed SDN technology aims at addressing the aforementioned challenges by separating the control and data planes. In the SDN paradigm, instead of designating the decision making to every active components in the network, this will be handled by a centralized controller called the \emph{network operating system (NOS)}. For instance, when a switch receives a packet, it chooses the proper action (forward, drop, modify, sending to the controller, etc.) based on the rules (flow table), which are defined by the programmable network applications at the centralized controller that rely on the NOS~\cite{Kreutz2015Software-DefinedSurvey}. The communication between the NOS and the forwarding layer or the data plane is established by some protocols such as OpenFlow~\cite{McKeown2008OpenFlow:Networks}. In contrast to the predecessor distributed architectures, this programmability leads to easy evolvable networks because the switches no longer need to interpret multiple protocols to make decisions individually. Rather, the network manager can manage and update the rules centrally. Moreover, OpenFlow offers a standardized interface that enables the integration of various heterogeneous devices from different vendors that can significantly simplify the operation of multi-vendor networks.

Since the emergence of SDN, most researchers have been focusing on wired networks. However, the emergence of next generation of mobile communication networks (5G) and IoT networks need effective resource allocation and interference management techniques which makes SDN a good paradigm to adopt for these wireless networks~\cite{Haque2016WirelessTaxonomy}. Before applying SDN to wireless networks, also called \emph{software defined wireless networking (SDWN)}, several challenges such as the nature of wireless channels, dynamic network topology, heterogeneity of devices, and shortage of resources need to be further investigated. However, the SDN-enabled wireless networks can offer key advantages for both users and providers because of their centralized network management approach. Some of the main advantages of the SDN model in infrastructure-based, non-infrastructure-based, and hybrid networks are summarized as followed~\cite{Haque2016WirelessTaxonomy,Mendonca2012TheEnvironments}.
\begin{itemize} 
\item \textit{Network slicing (network virtualization):} In general, providing different services through a single physical infrastructure is a challenging task. With SDN, the infrastructure provider can slice the physical infrastructure into distinct virtual networks to handle different services or providers~\cite{Haque2016WirelessTaxonomy}.
\item \textit{Effective traffic offloading:} A global view of the networks and a centralized management approach in SDN enables the service providers to offload the traffic on the right locations and devices in the network~\cite{Mendonca2012TheEnvironments}.
\item \textit{Intelligent routing:} The forwarding devices in the SDN-enabled networks send the status of the traffic load to the control layer, so that the controller can balance the traffic efficiently because it is aware of the traffic status of the other devices in the network~\cite{Mendonca2012TheEnvironments}.
\item \textit{Security enhancement:} The rules in the flow tables that are regularly updated by the controller allow the network operator to define and assign new security roles to every SDN-enabled nodes in the network in a response to the network status~\cite{Ding2014SoftwareNetworks}.
\end{itemize}

While SDWN can offer the aforementioned advantages, the nature of centralized and software-based control of the network can introduce new security threats to the system \cite{Schehlmann,Kloti,He2016SecuringNetworks,Kreutz2015Software-DefinedSurvey}. These threats are even more critical in large-scale and heterogeneous SDWN-based IoT networks. It is anticipated that, by $2020$, about $21$ billion things will be connected to the Internet \cite{Gartner2015}. These ubiquitous devices present specific security concerns due to their limited power, computing capability, and physical accessibility. Similar to other wireless networks, in an SDWN-based IoT network, the programmable devices can play a new role in packet forwarding, in addition to their traditional roles of sensing, monitoring and controlling \cite{Afghah_Springer,Afghah_CDC,Razi_TW}. As a result, the ramifications of attacks can intensify and propagate to the entire network very rapidly. Therefore, conventional security solutions such as transport layer security (TLS), secure sockets layer (SSL), and other public key infrastructure (PKI)-based protocols are no longer effective for large-scale IoT networks with billions of active devices \cite{Holz2012}. 

In this paper, we propose an anti-piracy protection mechanism based on development of resistive random access memory (ReRAM)-based PUFs to protect both the network and IoT hardware intellectual properties (IPs) from potential code-injection and rogue node injection attacks. A key advantage of PUF-based IP protection is that it can prevent the malicious entities from copying the hardware components by noting the unclonable properties of PUFs. While in common crypto-based solutions, an attacker can easily replicate the system components when it got access to one device, in a PUF-based method, even if an adversary has a physical access to some devices, cloning the intrinsic characteristics of the chips and emulating them to forge the identity of the things is very difficult or almost impossible. 
By using the embedded memory in devices to generate the PUFs, we can also reduce the cost of manufacturing sensors and devices, because keeping the secrets on the chips does not need nonvolatile memory (NVM) and an always-on power source \cite{Herder}.

\subsection{Contributions of the paper}
The key contributions of this paper are summarized as followed:
\begin{itemize}
\item A novel hardware security mechanism for IoT networks using ReRAM-based PUFs is proposed that utilizes the embedded memory of the devices.
\item Multi-state strong PUFs are developed, in which the natural drifts of PUFs' responses in different network conditions are predicted to reduce the error in challenge-response pairs of PUFs.
\item A new PUF-based PKI protocol is proposed to secure the controller in SDWN.
\end{itemize}

The rest of the paper is organized as follows: Section \ref{sec:SDWNarchitecture} provides an overview of the architecture of SDWN. We review recent security threats for SDWN and IoT, and the corresponding countermeasures in Section \ref{sec:background}. An introduction to PUFs and their applications in security is provided in Section \ref{sec:protocol}. We present the proposed security protocol for IoT networks using ReRAM-based PUFs with multi-state design. The experimental results are presented in Section \ref{sec:numerical} followed by some concluding remarks in Section \ref{Conclusions}. 

\begin{figure}

\centering
\includegraphics[width=4in]{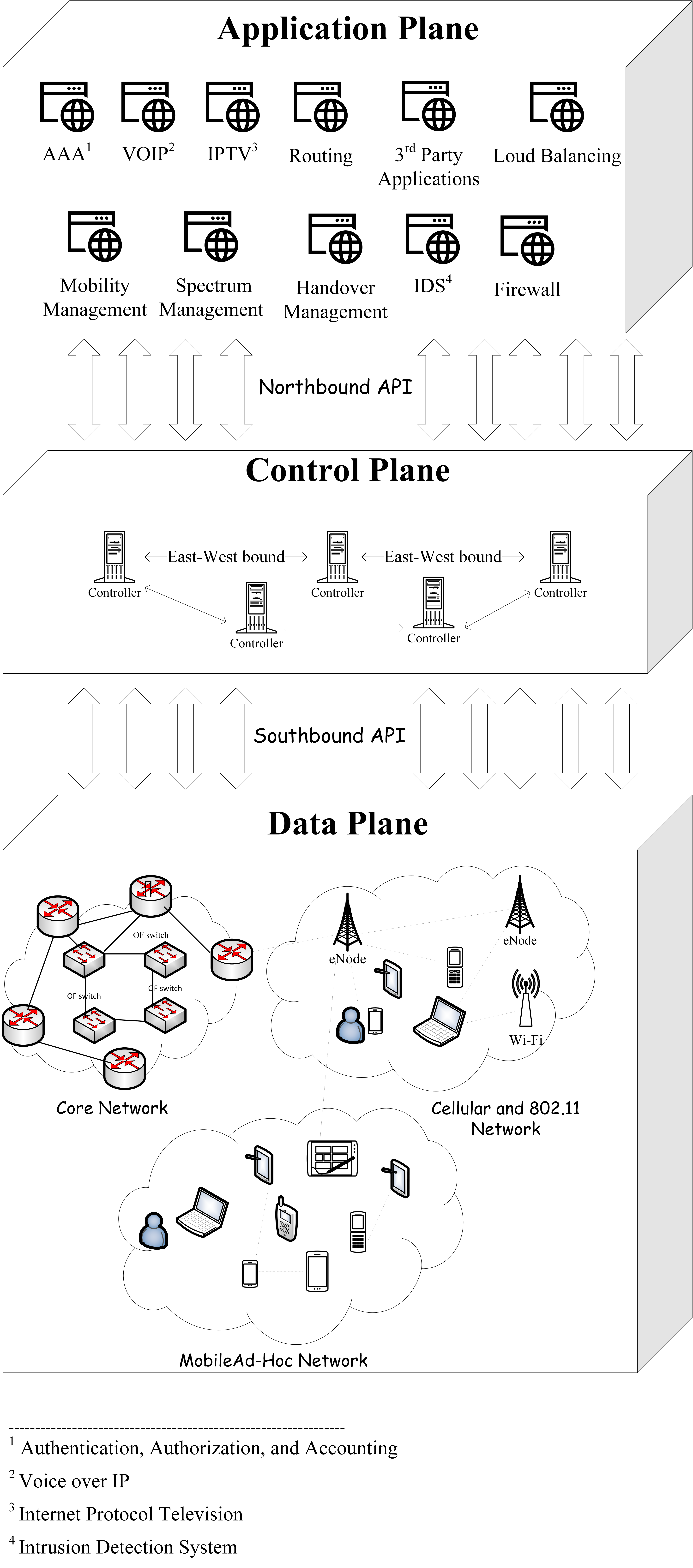}
\caption{General architecture of SDWN.}\label{SDWNarchitecture}
\end{figure}

\section{SDWN Architecture and Components}\label{sec:SDWNarchitecture}
As mentioned before, in the SDN paradigm the control plane makes decisions about traffic independent of the data plane which handles forwarding traffic to the target destinations. A general architecture for SDWN based on 3GPP evolved packet system is described in~\cite{Bernardos2014AnNetworking}. As shown in Figure 1, SDWNs consist of three layers:
\begin{enumerate}
\item \textit{Data Plane (infrastructure layer):} Recent wireless networks (infrastructure-based or infrastructure-less) have become increasingly heterogeneous. Hence, several heterogeneous radio access networks (RANs) such as Wi-Fi, 3G, LTE, LTE-A, and 5G can co-exist in this layer. In addition, the core networks (CNs) handle the communications among the users or servers. In mobile networks or wireless sensors networks, mobile terminals or wireless sensors also are included the data plane. 
\item \textit{Control Plane:} The NOS or mobile network SDN controller resides in this layer. The controllers receive rules and commands from the network applications and send them to the data forwarding devices in the data layer.  
\item \textit{Network Applications:} The operators or service providers use application program interfaces (APIs) to manage the controllers in the control plane. By using the network applications (e.g., mobility management, authentication, accounting, intrusion detection), the operators have access to network resources and manage them in an all-inclusive vision. 
\end{enumerate}

The communications interfaces among these three layers can be divided to three categories:
\begin{itemize}
\item \textit{Northbound Interface:} An interface to the operators, the service providers, or the application developers can be offered by NOS. The operators can dynamically manage the shared network resources between the virtual networks in one physical infrastructure. Furthermore, the service and application providers with different levels of access can influence on and change the network behavior using the authorized interfaces. 
\item \textit{Southbound Interface:} The way packets are forwarded between the forwarding devices is defined by the southbound application programming interfaces (APIs). An interface to the physical user plane network in the CN, RAN, or mobile node allows the network providers to set new policies and protocols or modify the old ones.
\item \textit{East-West bound Interface:} Although in SDWN, the controller is logically centralized, but to improve the scalability and robustness, it can be physically distributed. Therefore, east/westbound interface is responsible for data exchange among the distributed controllers.
\end{itemize}

\section{Background and Related Work} \label{sec:background}
In this section, we present a summary of state-of-the-art research on security issues related to IoT networks and software defined networks.
 
\subsection{Security Threats in IoT Networks}
The emergence of IoT networks raise a new set of security challenges due to the large number of devices connected to the Internet, the ad hoc nature of such networks as well as the energy constraints and limited computational capability at these devices \cite{ALABA201710,Diego,Zhao}. Moreover, the heterogeneity of IoT networks consist of various deceives with diverse set of security capabilities further threat the security of the network as a whole. Therefore, the existing cryptographic protocols that commonly involve intensive key generation, and key sharing process can not be utilized in such large scale networks. Several common security threats for IoT networks include: i) denial of service attack, ii) routing threats, iii) replay attacks, iv) fake node attack, v) side channel attack, vi) node capture, and vii) mass node authentication problem \cite{Zhao,Perrig}. As we describe in Section \ref{sec:protocol}, the proposed memory-based PUF technology can offer an affordable mechanism to protect the IoT networks from several of these attacks such as mass node authentication problem, and fake node attack.

\subsection{Security Threats and Countermeasures in SDWN} \label{sec:threats}
While the SDN paradigm can enhance the scalability and performance of traditional networks as well as strengthen the current security mechanisms because of its reprogrammable and reconfigurable properties, it can also introduce new security threats to the network\cite{Klingel,Jagadeesan,Zeadally}. Due to recent emergence of SDWN, there are a few reported work in the literature to study the security issues of SDWN. In~\cite{Chen2016Software-DefinedSecurity}, the authors reviewed and classified the security threats in software defined mobile networking (SDMN) such as spoofing, tampering, repudiation, information disclosure, denial of service, elevation known as STRIDE. The authors in \cite{Swapna} compare the performance of several communication protocols in SDWNs including border gateway protocol (BGP), NETCONF when facing STRIDE security threat. In \cite{He2016SecuringNetworks}, the authors provide a comprehensive survey of security threats of SDWNs related to a centralized controller as well as the separation of the control and data planes. 

Here, we present an overview of the security attacks and their countermeasures in SDN and SDWN ~\cite{Chen2016Software-DefinedSecurity,Kreutz2013TowardsNetworks,Ali2015ANetworking,He2016SecuringNetworks,Jagadeesan}. 

\subsubsection{Data Plane Security Issues}
In attacks regarding the data plane, the attackers can target different network elements including the forwarding components such as the OpenFlow switches, the mobile terminals or the sensors in IoT or the radio access elements. The various threats on the forwarding components can be classified as follows:
\begin{itemize}
\item Sniffing: a passive attacker can sniff the traffic at the forwarding components for use in future attacks. 
\item Forged traffic flows: the compromised nodes or even malfunctioning or mis-configured devices can generate forged traffic flow and send it to the controller. An active attacker can force to send false flow information to the controller to deceive it. 
\item Flow tables overflow: an attacker can continuously send a series of flows which are slightly different from each other. As a result, the switch is forced to send the information of the new flows to the controller and receive related entries updates. Since the flow table on the OpenFlow-enabled devices can only accept a limited number of entries, this process can cause the flow table to become full very quickly and slow down that impacts the forwarding of regular flows~\cite{Chen2016Software-DefinedSecurity}.
\item Selfishness: some mobile nodes in mobile ad-hoc networks (MANET)s or wireless sensor networks (WSNs) may tend to act selfishly to preserve their resources (such as power and computing). As a result they may refuse to forward and relay the packets received from other users or devices. 
\item Spoofing (forged identity of authorized components): in this threat, the attackers want to conceal their malicious activities behind a legitimate entity in the network. Hence they try to forge the identity of authorized terminals or switches to achieve their goals. The infrastructure-less environment exacerbates the frequency of this attack due to the easy access to the communication media. 
\end{itemize}
 
Several methods have been proposed to mitigate these attacks including i) encryption of data and sensitive parameters at the forwarding elements which is the first line of defense against adversaries to sniff or steal credential data, ii) using intrusion detection systems (IDSs) or intrusion prevention systems (IPSs) to detect or prevent anomalous behaviors of the components, iii) mutual authentication between the controller and the forwarding components which can prevent the unauthorized access as well as forging the identity of legitimate users.

In addition to the forwarding components' threats, the RANs can also suffer from a wide variety of attacks such as sniffing and denial of service attacks (DoS), mainly due to the co-existence of multiple standards and mobile networking technologies at the data plane layer. These threats can be thwarted by different encryption and mutual authentication mechanisms.

\subsubsection{Control Plane Security Threats}
The centralized nature of network management in SDN-enabled networks make them prone to several attacks that threaten the centralized controller. Hence, failure of the centralized controller called as "single point of failure" has been the main concern in these networks \cite{He2016SecuringNetworks}. The implementation of a distributed controllers' architecture for SDN-enabled IoT mitigates the risk of single point of failure and also enhances the security level of the network by using hierarchical controllers in multiple domains~\cite{Flauzac2015SDNSecurity}. Other major threats to the control plane include: i) Distributed denial of service (DDoS) attacks or flood attacks, in which the control plane is overwhelmed by dummy flow traffic from the attackers and it has to respond to these unknown flows for making decisions, and ii) Data leakage, where an adversary can discover the policy of the controller about special flows by using packet processing timing analysis~\cite{Scott-Hayward2016ANetworks}.

\subsubsection{Application Plane Security Issues} 
If an attacker takes over control of the clients remotely or physically by means of viruses, trojans, and etc., the attacker can insert fraudulent flow rules into the forwarding components and potentially control the network. To prevent these threats, the admin terminals can be protected by traditional mechanisms such as anti-virus and IDS. Moreover, the likelihood of these attacks can be reduced by using two-factor authentication mechanisms when accessing the applications and the clients as well as by choosing strong access control policies. In addition to these client-based threats, the network application threats refer to attacks that implement network functions in applications running on the application plane which can potentially disturb the performance of the entire network~\cite{Ahmad2015SecuritySurvey}. Regular penetration testing and strong authentication(e.g. Kerberos \cite{Neuman}) and authorization management techniques can prevent unauthorized access by applications executed by the controllers \cite{He2016SecuringNetworks,Scott-Hayward2016ANetworks}.

\subsubsection{Communication Protocols Security Issues}
The attacker can exploit the vulnerabilities in the protocols that control the communication between the controllers, network applications, switches, base stations, and users’ devices to launch DoS attacks on the entire network or sniff important information. In ~\cite{Chen2016Software-DefinedSecurity}, the authors presented three well-known threats and discussed countermeasures to mitigate them:
\begin{itemize}
\item IP spoofing: the lack of the IP layer security (such as internet protocol security (IPsec)) among backhaul of RANs can lead to this attack.
\item Transport layer security/secure socket layer (TLS/SSL) vulnerabilities: recently many flaws have been found in these security protocols such as SYN DoS that can be launched by the attackers.
\item Man-in-the-middle attack: in this attack, the adversary intercepts the communication channel and exchanges the authorized parties’ messages in a way that they are not aware of the existence of any adversary. 
\end{itemize}

Mutual authentication mechanisms together with the key distribution algorithms can be applied for securing the channel. Additionally, we can use improved protocols such as the host identity protocol (HIP) and IPsec tunneling to secure the channels between the controllers and the forwarding components ~\cite{Chen2016Software-DefinedSecurity,Liyanage2015SecurityNetworks}.

\section{Proposed Security Protocol using Memory-based Physical Unclonable Functions (PUFs)} \label{sec:protocol}
PUF is a generic technology used for creating cryptographic primitives that can be integrated in cyber physical systems (CPS) to strengthen security \cite{Maes,Maes_Springer,Jin,Herder,Pappu}. The concept was introduced 15 years ago, and has been commercialized quite successfully recently. During manufacturing, electric components encounter random variations that are due to small local changes in the chemical composition, physical dimensions, density, and other physical elements \cite{Cambou_patent,Gao,Guajardo}. These variations make each device unique. The idea behind a PUF is to identify these differences in order to be able to differentiate each component from the others so that we can achieve a secure authentication of the components in the CPS. 

The basic protocol is initiated by generating PUF ''challenges'', the reference patterns of the components that can act as digital finger prints. These challenges are usually stored in a secure server for future use. When queried by the secure server, which is the case during an authentication cycle, the PUF generates ''responses'' in a way similar to the challenges generated upfront. The authentication is thus, completed by analyzing the challenge-response-pairs (CRPs), and the resulting matching error rate. 
This methodology is not different from what is done to authenticate users in biometric methods using their finger-prints, images of their iris, veins, or biometric characteristics. Two important figures of merit for PUFs include: i) the ability to be clearly identifiable in spite of natural drifts, or noisy conditions, and ii) the existence of secret properties that make them hard to extract through side channel analysis often used by the hackers. These two benefits are often in conflict with each other, and fuzzy PUFs that are hard to extract by the hackers could also be erratic under noisy conditions.

In this work, we utilize the embedded ReRAM in IoT devices to generate strong PUFs. Since these elements operate at very low voltage and low power, they are hard to analyze through side channel attacks. We also propose a multi-state and machine learning based technique that greatly strengthens the PUFs by reducing CRP error rates. 
This involves developing novel hardware design and computational mechanisms to create a PUF with multi-state memory, where the measurements of a physical parameter are saved in multiple states format rather than the conventional binary style \cite{Cambou_Orlowski}. To do so, we propose a novel design for the CRP generation process that captures the specific ''personality'' of the physical elements underlying the PUFs under various conditions (such as ambient temperature). This can substantially improve the accuracy of the challenge and response evaluation and hence reduce the error in the challenge and response comparison.

\subsection{Memory PUFs Compared with Legacy PUFs, and Why ReRAM?}
The early PUF technology was based on ring oscillators and gate delays \cite{Tuyls,Suh}. The authentication protocol of such PUFs uses the result of ''in situ'' matching of the challenges with the responses. The secure server sends the challenge to the PUF, the PUF responds with a positive or negative authentication of whether the frequency or delay matches or not. Such a protocol is very interesting because there is no need for complex key distribution protocols, as there are no keys stored on the PUF, therefore the crypto-analyst cannot easily find the cryptographic primitives, and the authentication process can be quick.

PUF technology is not easy to achieve in practice however, because the physical elements can vary when subject to temperature changes, parameter drifts, bias effects, electromagnetic interferences, and aging \cite{Helfmeier,Potkonjak,Maiti}. Drifting responses produce higher CRP error rates, and can create false negative authentications. The attackers, through side channel analysis and fault injection, can extract the responses from PUFs defeating their purpose.

Memory-based PUFs are now becoming increasingly important as a cryptographic primitive to protect IoT devices \cite{Maes_Tuyls,Holcomb,Christensen,Zhu,Krutzik,Chen,Vatajelu,Cambou_IoT}. Embedded memories are widely available in IoT devices as cache memory, or non-volatile storage. The density needed for PUF CRP generation is extremely small, i.e. 128 to 256 bits, compared with the memory needed in IoT devices, which is typically in the 1-8 Mbit range. The low memory requirement of PUFs can be easily achieved by IoT devices making them hard to extract through side channel attacks.   

The only operations available with memory devices are: program, erase, and read. Hence it is not possible to follow the protocol described above with ring oscillators and gate delays to match ''in situ'' challenges and responses. During authentication cycles, it is then necessary to extract the responses away from the memory array, and to perform the CRP matching separately. Two protocols emerge, the first is to send the response to the secure server which analyzes the CRPs matching. In the second method, the analysis of the CRP matching is directly performed in the IoT device.
 In both cases the communication between the secure server and the IoT device has to be encrypted to protect the PUF’s cryptographic primitives, challenges or responses. For this purpose, a crypto-processor has to be incorporated as part of the design of the IoT device, and cryptographic protocols such as public key infrastructure (PKI) need to be in place. The deployment of PKI requires the distribution of private keys to the IoT devices. These keys can be stored in the embedded memory. 

ReRAM is an emerging technology for IoT that has the potential to replace EEPROM and flash as a non-volatile memory \cite{Ghosh,Gilbert,Kozicki_5,Kozicki_6,Kozicki_11,Kozicki_4,Mickel}. ReRAMs operate at very low power compared with flash, have low access time, and are very fast to program. These properties are extremely desirable for secure operations.
Differential power analysis (DPA), and electro-magnetic interference analysis coupled with fault injections are not effective in extracting the secret keys that are stored in ReRAMs. This is because their operating power is orders of magnitude lower than flash. Hence, ReRAMs operate below the noise level present in the system \cite{Yoshimoto,Cambou_IoT,Cambou_Afghah,Cambou_Orlowski,Afghah_patent}.
Electron beams created by secondary electron microscopy (SEM) can be deflected by the electrons trapped in flash memory thereby exposing the content of the stored information. In contrast, the chargeless ReRAMs are immune to this type of attack. Therefore, both metal-oxide ReRAM, and conductive bridge ReRAM technologies are appropriate candidates for the purpose of PUF CRP generation in IoT networks \cite{Cambou_IoT,Kozicki_05,Cambou_Afghah,Cambou_Orlowski}.

\subsection{Proposed Memory-based PUFs with Multi-state and Machine Learning}
\begin{figure}[tb]
    \centering
    \includegraphics[width=0.8\textwidth]{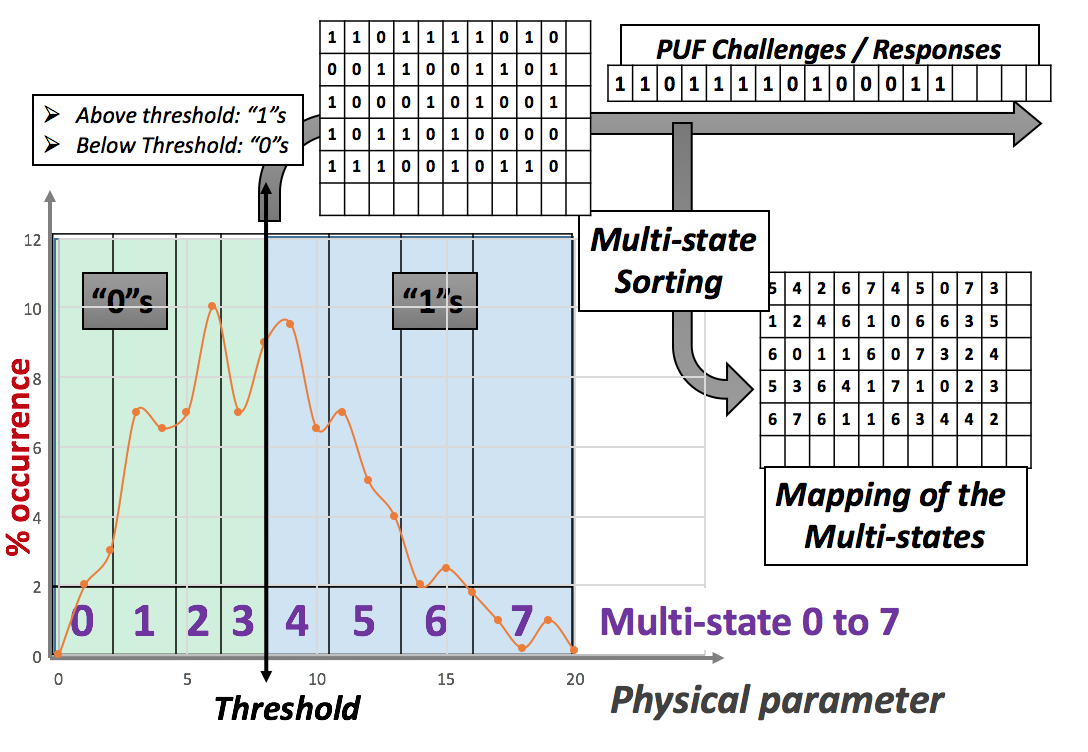}
    \caption{challenge/response generation in the proposed PUFs with multi-states}
    \label{multi-state}
\end{figure}

In the proposed multi-state PUF design, the challenges or responses are generated based on the measurements of a physical parameter such as temperature or bias voltage, $V_{\text{set}}$, and are saved in multiple states format rather than the conventional binary style, as depicted in depicted in Figure \ref{multi-state}. In the conventional binary notation, a $''0''$ refers to the case where the measured parameter of a memory cell is below the threshold located in the middle of the distribution, while a $''1''$ is programmed in the cells measured above the threshold.

In our proposed multi-state method, the cells are organized in $n$ \emph{multiple states} by sorting out the value of the physical parameters underlining each cell of the memory, as shown in Figure \ref{multi-state}. This multi-state method can more accurately capture the specific ''personality'' of the physical elements underlying the PUFs in a challenge or response generation process. A PUF of $N$ bits is to be sorted into $n$ states, either during challenge generation, or response generation. Each state $i$ has $n_i$ cells such that $\sum_{i=0}^{n} n_i=N$. 
The PUF responses are generated in the same way as the challenges as often as needed.  CRPs errors are to be expected because the measurement of the physical parameters of the PUFs is changing over time. 

\begin{figure}[h]
    \centering
    \includegraphics[width=0.8\textwidth]{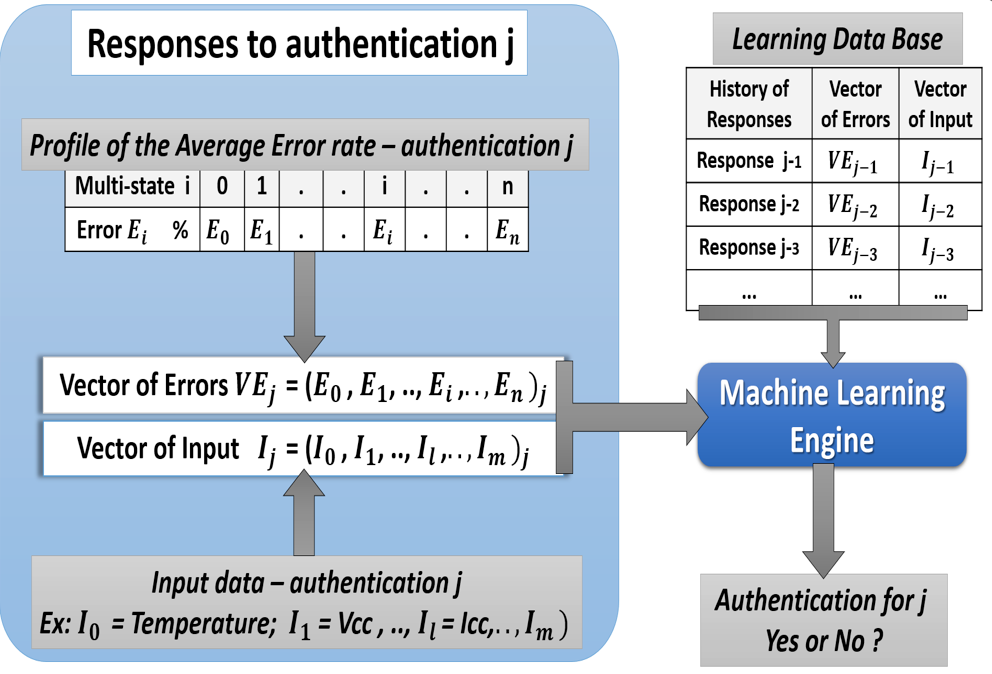}
    \caption{ CRPs error rates in multi-state PUFs}
    \label{CRP}
\end{figure}

For a given cell $k$ that is part of the PUF, the CRP error between the challenge $C_k$ and the response $R_k$ is given by $\Delta CRP_k=|R_k-C_k|$, where $\Delta CRP_k$ is the CRP error rate of the cell. For the populations of $n_i$ cells that are part of the state $i$, the average $CRP$ error rate is given by:
\begin{align} \label{eq:1}
E_i=\frac{1}{n_i} \sum_{k=1}^{k=n_i}|R_k-C_k|
\end{align}
The average error rates $E_0$ to $E_n$ (as calculated with (\ref{eq:1})) for the $n$ states result in a Vector of Errors (VE) that is characteristic of a particular response:  $VE=(E_0, E_1,...,E_i,...,E_n)$. This process is summarized in Figure \ref{CRP}.

\begin{figure}[hb]
    \centering
    \includegraphics[width=0.8\textwidth]{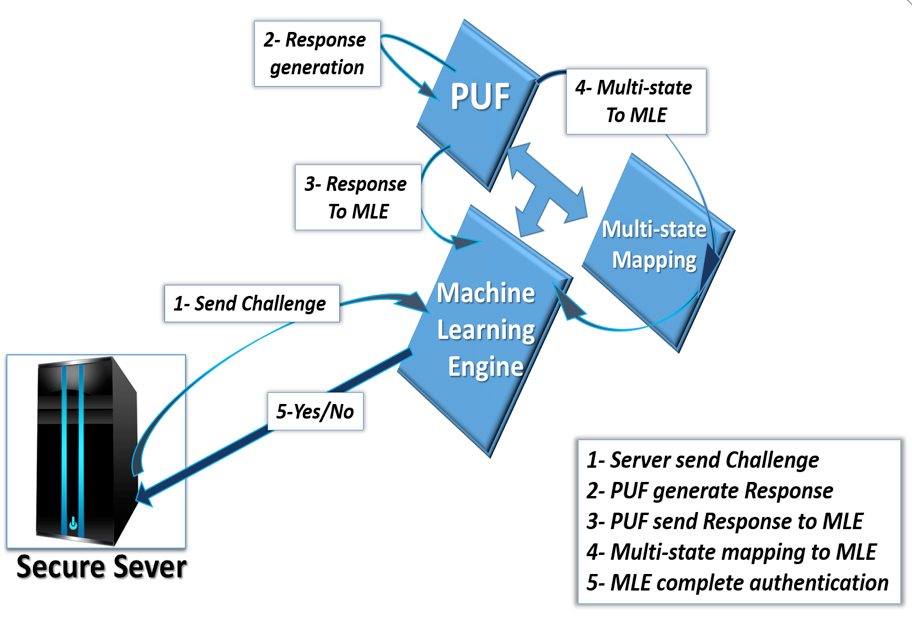}
    \caption{Authentication protocol for memory-based PUFs with a machine learning engine (MLE)}
    \label{diagram}
\end{figure}

These VEs are used to complete the authentication process using a machine learning engine (MLE) that predicts the expected drifts of the responses for a given physical parameter (such as temperature), and adjust the results accordingly. When the server sends a challenge to the MLE, a fresh response is generated by the PUF. The MLE gathers the response, as well as all available data to compute a secure authentication. The MLE integrated in the micro-controller handles the communication between the secure server and the PUF. For the authentication $j$, $VE_j=(E_0, E_1,...,E_i,...,E_n)_j$ and the vector of input $I_j=(I_0, I_1,...,I_i,...,I_m)_j$, where $m$ denotes the number of input parameters, are fed into the MLE. The vector of input includes the physical parameters of interest such as operating temperature, and biased conditions. Then the MLE completes the authentication process by considering the available learning data based on a record of prior responses with the predictive models of the laws for the PUF parameters. It is worth mentioning that noting the limited size of input history dataset, this process does not impose a considerable computational load to the IoT devices. The block diagram of this authentication protocol is shown in Figure \ref{diagram}.

\subsection{Proposed Security Protocols for SDWN-based IoT Networks}
\begin{figure}[h]
\centering
\includegraphics[width=5in]{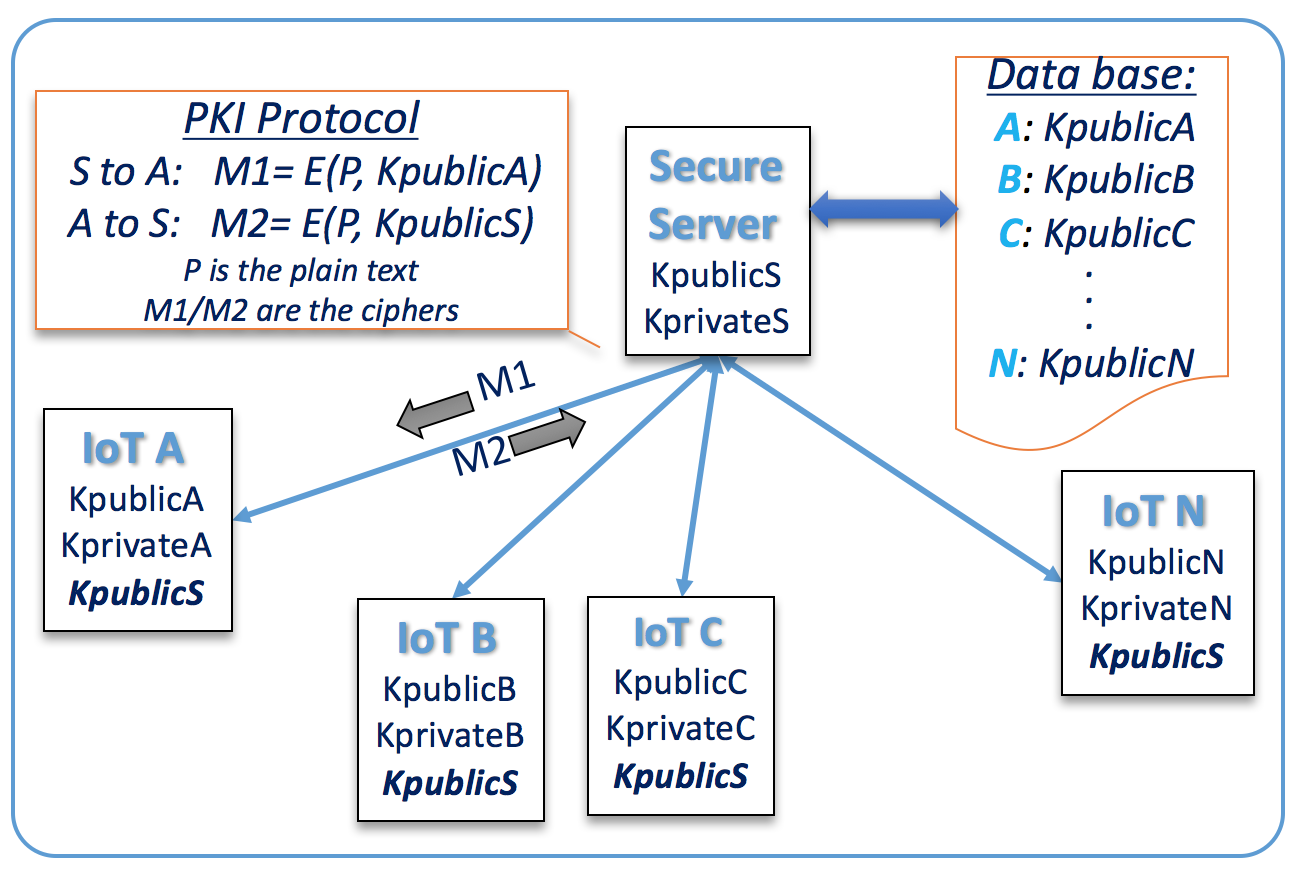}
\caption{Block diagram of a public key infrastructure protocol to secure the network with IoTs.
The two-way communications between the server and the IoT are encrypted with public keys.}
\label{PKI}
\end{figure}

We propose a security protocol based on the developed memory-based PUFs to significantly enhance the security in SDWN-based IoT networks. Public key infrastructure (PKI) is known to be a powerful and commonly-used infrastructure to protect software defined wireless networks with a large quantity of IoT devices, and peripherals. When the PKI security protocol is utilized in SWDNs, each node needs to have a pair of public and private keys to allow two-way encrypted communication between the IoT devices and the secure controller as depicted in Figure \ref{PKI}.  
The private keys can be downloaded during the post manufacturing operations of secure elements; these operations are also called “personalization”. If the non-volatile memory of the secure element is made with ReRAM rather than flash, the private keys can be adequately protected from an attack. 

However, one drawback of this method is the public-private key pair of the secure controller.  Attackers can focus their efforts to break this single node, and compromise the entire group of IoT devices connected to the controller. Another threat, although with potentially limited impact, is the loss of a public-private key pair of one IoT device to some third party. 

\begin{figure} [h]
\centering
\includegraphics[width=5in]{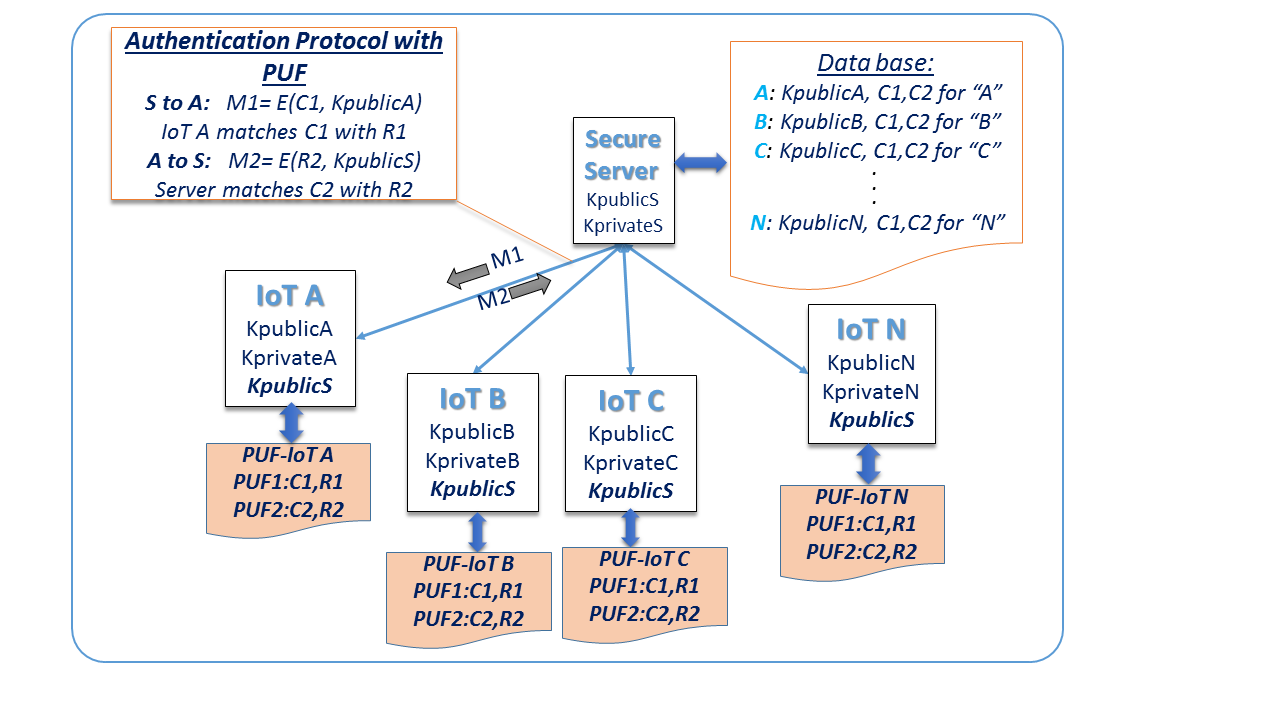}
\caption{Block diagram of the proposed PKI protocol with PUF authentication.
The two-way communication is also encrypted with public keys. Two PUFs per IoT device provide  two-way authentication.}
\label{PUF_protocol}
\end{figure}

Here, we propose a novel PUF-based protocol that can drastically reduce the exposure to these cyber-attacks.
In this protocol, two challenges of $C1$ and $C2$ are generated from two distinct parts of the array at every IoT node in the network, and these challenges are stored in the secure network. We describe the protocol below. 
\begin{enumerate}
\item The first step of the protocol is initiated by the secure server; an encrypted challenge $C1$ is sent to the corresponding IoT device. Then IoT device decrypts $C1$, generates a response $R1$ from the part of the array that generated $C1$, then checks whether the CRPs match. This step authenticates the secure network. 
\item The second step of the protocol is initiated by the IoT device; a response $R2$ is generated from the part of the array that generated $C2$, $R2$ is then encrypted and sent to the secure server. The secure server compares $C2$, and $R2$ to authenticate the IoT device. 
\end{enumerate}
Figure \ref{PUF_protocol} presents a block diagram of the proposed PUF-based PKI protocol.

If an attacker breaks the public-private key pair of the server, it should not be able to authenticate itself by the IoT device, unless the hacker also finds a way to uncover $C1$. If this were to happen, other IoT devices would not be exposed to the breach, thereby protecting the network from a large scale attack. Conversely, if an attacker breaks the public-private key of the IoT device, it should not be able to be authenticated by the secure server, unless the attacker also finds a way to uncover $C2$. Our proposed PUF protocol therefore provides an important second level of protection in addition to the protection offered by PKI. 
The protocol presented in this section can be extended to a larger number of PUFs by a memory array. This protocol could be used for hierarchical level of security, with additional PUFs needed for highly sensitive parts of the network.

\section{Experimental Results} \label{sec:numerical}
\begin{figure} [h]
\centering
\includegraphics[width=4in]{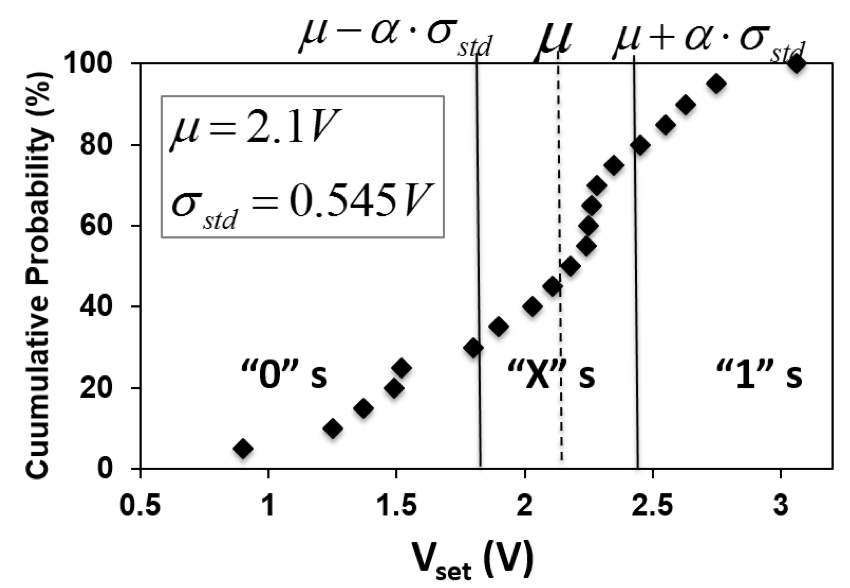}
\caption{Cumulative $V_{\text{set}}$ probability distribution for the entire array of cells.}
\label{exp1}
\end{figure}

\begin{figure} [h]
\centering
\includegraphics[width=4in]{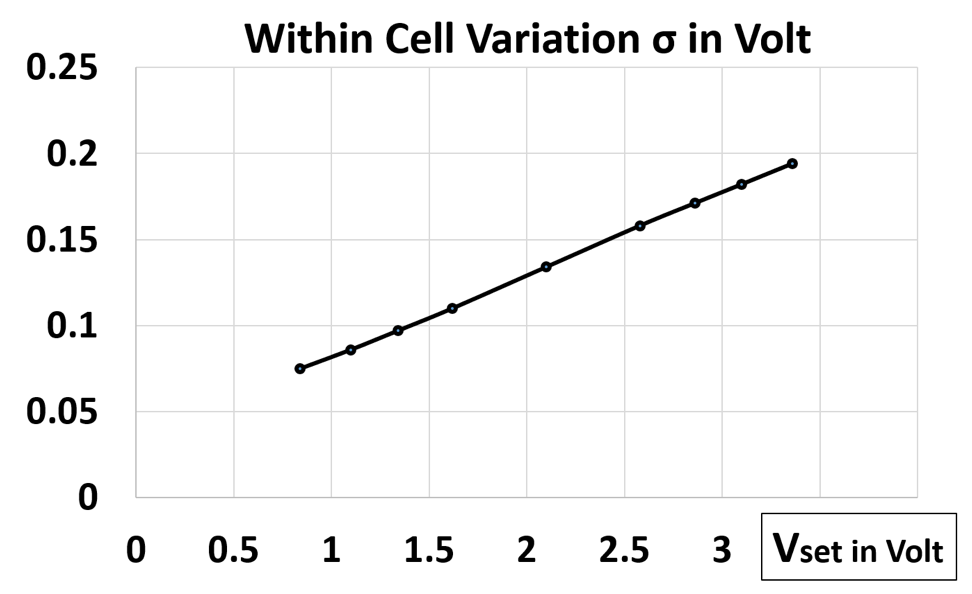}
\caption{Variation of $\sigma$ within cell versus the average $V_{\text{set}}$}
\label{exp2}
\end{figure}

In this section, we present the experimental results obtained to generate PUFs using ReRAM based on metal oxide with oxygen vacancies is presented. For this purpose, Cu/TaOx/Pt resistive devices have been fabricated, and characterized at Virginia Tech in a crossbar array on a thermally oxidized silicon wafer \cite{Liu,Liu2,Liu3,Ghosh}. 
Figure \ref{exp1} shows the cumulative $V_{\text{set}}$ probability distribution within a typical sample of ReRAM memory array, containing a large number of cells. The mean of this distribution is $\mu=2.1$V, as indicated by the dashed line, and the standard deviation is $\sigma=0.54$ V. The variation of standard deviation ($\sigma$) of all cells are extrapolated as shown in Figure \ref{exp2}.

The challenges and responses on each cell of the array are generated based on the value of $V_{\text{set}}$, the voltage necessary to create a conductive bridge within the solid electrolyte separating two conductive electrodes. In order to study the quality of the PUF, the samples were submitted to repetitive program erase cycles.  When a progressive positive voltage sweep is applied to the cell, the programming step, the voltage reaches $V_{\text{set}}$ when a conductive filament is created, which reduces the resistance by two or three orders of magnitude. With a negative voltage sweep, the erase step, has the reverse effect; when the voltage exceeds $V_{\text{reset}}$ the conductive filament is partially dissolved, and the resistance increases by several orders of magnitude. Based on the experimental data presented in Figure \ref{exp1}, the drift between the challenges and the responses is modeled by using a normal distribution.

Here, we applied a \emph{ternary states} methodology, with a threshold value $T$ which is close to the median value of the $V_{\text{set}}$. When a cell has a $V_{\text{set}}$ clearly below $T$, the cell is considered as a ''0'' state; when a cell has a $V_{\text{set}}$ clearly above $T$, the cell is considered as a ''1'' state; and when a cell has a $V_{\text{set}}$ close to $T$ or one that varies randomly around $T$, the cell is considered as a ''X'' state. As a result, the challenge-response-pair CRP error rate is reduced when the proportion of X states is higher, leaving only solid "0" and "1" states.

One metric to evaluate the performance of PUFs is the amount of entropy, and therefore the number of independent CRPs that can be generated which is limited by the number of elements used to construct the PUFs. Because the traditional PUF generation mechanisms have certain limitations in this regard, several mapping methods have been proposed to produce stronger PUFs. In \cite{Maiti}, the authors proposed an identity-mapping function to expand the set of challenge-response pairs for ring-oscillator PUFs, where a group of ring-oscillator frequencies is utilized to generate the PUFs. Because generating stronger PUFs of this type comes with higher area cost, this proposed method can generate stronger ring-oscillator PUFs with lower cost using an identity-mapping function that results in a larger set of CRPs. While the new sets of CRPs are not information-theoretic independent, the statistical tests confirmed that the generated lower-cost PUFs with the identity-mapping function are strong.
Noting the fact that the PUFs usually need only 128 to 256 bits to ensure an acceptable level of security in different applications, while commercial memory arrays that are integrated within micro-controllers, ordinarily have memory densities in the mega-byte range, we can easily generate a large set of CRP for our proposed memory-based PUFs. 
As shown in figures \ref{exp1} and \ref{exp2}, this value follows a normal distribution with mean value of $\mu$, where its standard deviation varies cell to cell. This concludes the uniformity of the responses. Furthermore, considering the scale of available memory, the readings can be done over different cells in a way that there is no overlap between the challenges (or the responses) that confirms the inter-response dependency. 

Moreover, we enhance the reliability of the developed PUFs by predicting their response variations for different environmental conditions (e.g. temperature) as presented in Figures \ref{exp1} and \ref{exp2}.
In memory-based PUFs, this criteria means to have enough random variations, cell to cell, in order to obtain strong cryptographic entropy, while the measurements of each cell should be reproducible when responding to successive queries. This can be satisfied when the standard deviation of cell to cell (mainly due to manufacturing variations) is much higher than the standard deviation of each cell (mainly due to noise, and measurement variations).

In this experimental validation we also utilized the aforementioned learning approach to predict the natural drifts in the responses of a PUF in different situations. The $V_{\text{set}}$ of ReRAMs is sensitive to temperature, and biased conditions. When the temperature increases, the mobility of the positive ions, oxygen vacancies, is higher, and conductive filaments are created at lower voltages, hence the $V_{\text{set}}$ is lower. If the generation of the challenges, and responses is done under different conditions this could increase CRP error rates. Considering this fact, the drifts that are due to temperature changes, or different biased conditions are largely predictable by the laws of physics and are tracked by the learning approach.

\begin{figure} [h]
\centering
\includegraphics[width=4in]{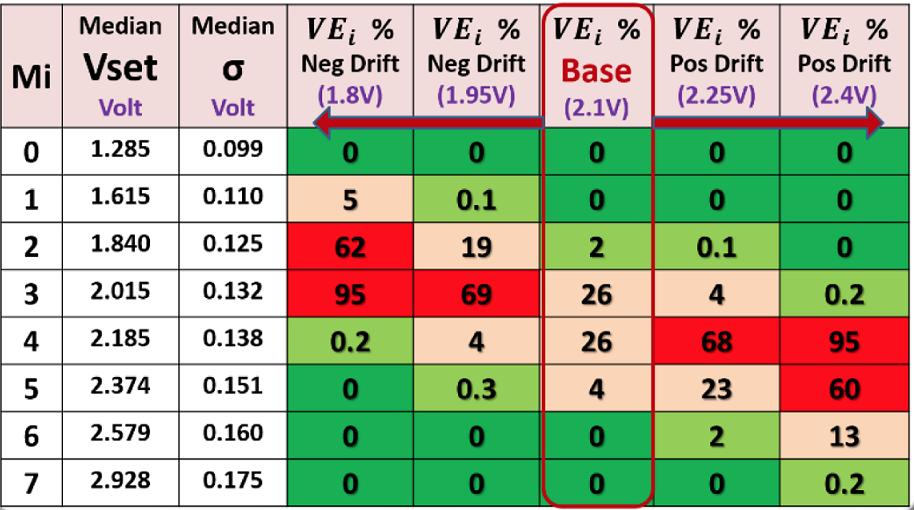}
\caption{Statistical analysis for the impact of the drift of the responses on the CRP error rate.}
\label{exp4}
\end{figure}

\begin{figure} [h]
\centering
\includegraphics[width=4in]{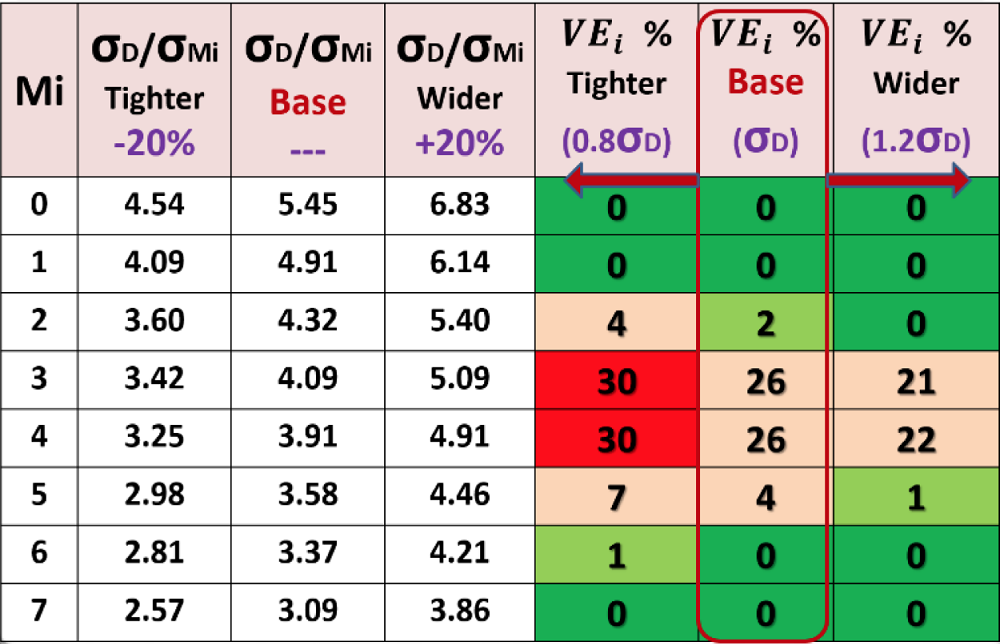}
\caption{Variations of standard deviation for the entire population versus the standard deviation of each cell.}
\label{exp5}
\end{figure}

An analysis of the results shown in figure \ref{exp4} reveals, the impact of the drift of the response on the CRP error rates, by state. In this Figure $M_i$ denotes the state $i$ for the proposed multi-state PUF. The population of the ReRAM array with 8 multi-states of 0 to 7 with equal probability of 12.5\% has been considered. The $VE_i$ vectors are calculated by state from 0 to 7 for the base, where the resulting base vector of error has a mean of $V_{\text{set}}$ of 2.1 Volt. 
When the responses drift in a positive direction, respectively to 2.25V and 2.4V, the CRP error rates of the first four states decrease, while the CRP error rates of the last four states increase. The effects are reversed for negative drifts (1.95V \& 1.8V). In Figure \ref{exp5}, the analysis is related to the respective change of the standard deviation of the entire population versus the standard variation of each cells. In this figure, $\sigma_D$ denotes the standard deviation of the entire distribution for all cells, and $\sigma_{M_i}$ is the standard variation of the distribution of all cells with state $M_i$.
If the spread of the general population of responses to the PUF is getting tighter compared with the spread of responses to an individual cell, the average error rates across the 8 states will go up. Conversely, if the spread is relatively wider, the average defect rates will decrease. 

This proposed method to capture the profile of the physical parameters underlying a PUF with multi-states can result in a tracking of the PUFs’ drifts over time that are predictable. As we mentioned earlier, the modeling of the effect of the external parameters such as temperature and bias conditions can decrease the probability of false negatives when authenticating a PUF-based IoT device under various conditions.

%

\section{Conclusions} \label{Conclusions}
One of the key challenges facing implementation of IoT networks is security. This is even more critical in SDWN-based IoT network noting the vulnerability of the network to the central controller failure due to malicious attacks. In this paper, we propose a novel ReRAM-based PUFs that can function as digital fingerprints to secure SDWN-based IoT networks. In order to enhance the performance of these PUFs in terms of reducing the error rate between the challenge and response pairs (CRPs) in different network condition, we proposed a multi-state machine learning technique. In this method, the potential drifts in the PUFs' responses due to various physical parameters such as temperature, and biased conditions are predicted and utilized to reduce the CRP error rates. The effectiveness of this method in reducing the CRP errors is confirmed in the numerical results. Furthermore, we proposed a PUF-based PKI protocol to establish a two-way authentication in SDWN-based IoT networks that protects both the server and IoT devices. This method adds another level of security comparing with common PKI protocols in a way that the attackers cannot authenticate themselves in the network by finding the public-private key pair, unless they can get access to the challenge. This can significantly enhance the security of the network specifically against the central controller attacks, because even if an attacker breaks the public-private key pair and the challenge of the server for one IoT device, other IoT devices would not be exposed to this attack, thereby protecting the network from a large scale attack. 

\begin{acknowledgements}
This project has been partially supported by Arizona Board of Regents under grants 1003074 and 1003074. The authors would like to thank their colleagues from the pilot manufacturing facility at Virginia Tech that allowed us to produce quality samples for this work. 
We thank the anonymous reviewers for their valuable comments which helped us improve the quality and presentation of this paper.
\end{acknowledgements}


\bibliographystyle{plain}

\end{document}